\documentclass[%
 reprint,
 amsmath,amssymb,
 prd
]{revtex4-2}

\usepackage{physics}
\usepackage{graphicx}
\usepackage{color}

\newcommand{\NSIDE}{N_{\mathrm{side}}}

\begin{document}

\title{Statistics of  non-polarized points in the CMB polarization maps}

\author{Jaan Kasak}
\email{jaan.kasak@abzu.ai}
\affiliation{Niels Bohr Institute, University of Copenhagen, Blegdamsvej 17, DK-2100 Copenhagen, Denmark}

\author{James Creswell}
\email{james.creswell@nbi.ku.dk}
\affiliation{Niels Bohr Institute, University of Copenhagen, Blegdamsvej 17, DK-2100 Copenhagen, Denmark}

\author{Hao Liu}
\email{ustc\_liuhao@163.com}
\affiliation{School of Physics and Material Science, Anhui University, 111 Jiulong Road, Hefei, Anhui, China 230601}

\author{Pavel Naselsky}
\email{naselsky@nbi.dk}
\affiliation{Niels Bohr Institute, University of Copenhagen, Blegdamsvej 17, DK-2100 Copenhagen, Denmark}

\begin{abstract}
    The non-polarized points (NPP) of the $Q$ and $U$ Stokes parameters of the CMB can be classified according to the geometry of the polarization field. We describe a procedure to identify these points in the pixelized sky and present the shape of the polarization angles in the vicinity of NPPs. We design a test of Gaussianity using the Kullback-Leibler divergence. We show that the total number density of non-polarized points of the E- and B-families is closely related to the presence of lensing and the tensor-to-scalar ratio $r$. We further show that in the absence of lensing, the total number of NPPs of all types does not depend on $r$, while the lensing effect removes this degeneracy. This analysis is applied to the CMB maps from the 2018 Planck release. We show that there is general consistency of SMICA and NILC maps compared to a reference set of Gaussian simulations. The strongest discrepancies are found in the Commander (with corresponding $p$-value $0.07$) and NILC ($p = 0.15$) maps.
\end{abstract}

\maketitle

\section{Introduction}
\label{sec:intro}

Verifying the Gaussianity and statistical isotropy of the CMB polarization is an important test of the Lambda-CDM model as well as inflation and other early universe physics.
In this direction, taking under consideration forthcoming CMB experiments, statistical characterization of the B-mode of polarization is especially important \cite{seljak1997,zaldarriaga1997,kamionkowski1997}.
Much effort has been made towards accurate determination of power spectra, and in addition other estimators and techniques have been developed to characterize the polarization fluctuations.
These include Minkowski functionals, peak analysis, power asymmetry, parity asymmetry, single- and multi-dimensional moments, etc.\ (see for review \cite{planck2019a,planck2019b}).
Since the anomalies of the CMB up to now do not have theoretical bases, it seems to be very important to investigate all possible estimators of non-Gaussianity and statistical anisotropy in order to split the effects of systematics and component separation (foregrounds) from primordial sources. 

In this direction we would like to draw attention to the statistical peculiarities of the $Q$ and $U$ components of the Stokes ``vector'', related to the non-polarized points (NPP) in the intensity map $I = \sqrt{Q^2 + U^2}$ and the corresponding features of the polarization angle $\tan(2\Psi) = \frac{U}{Q}$.
In principal, the Stokes parameters $Q$ and $U$ comprise a continuous spin-2 valued function on the sphere.
Because of their continuity, there will be well-defined contours along which $Q = 0$ and similarly contours along which $U = 0$ (so called contours of percolation \cite{1995ApJ...444L...1N}).
At the points where these contours intersect, $Q$ and $U$ are simultaneously 0, and the total sky signal is unpolarized \cite{naselsky1998}.
At the same time, these NPP manifest themselves as points of absolute minimum (zeros) of the polarization intensity, which for Gaussian $Q$ and $U$ has a Rayleigh distribution.
Thus, there is a connection between the statistics of NPP and the statistics of minima of the non-Gaussian field of polarization intensity.
Note that the existence of these points is a natural product for correlated pseudo-vector field as points of connection of the domains with different polarization, similar to the domains in ferromagnetics.
The number density of these points, their morphology, and their relative concentration will reveal some peculiarities of the morphology of the polarization due to different systematic effects, instrumental noise, and residuals of the component separation technique. 

In \cite{naselsky1998}, these non-polarized points for random Gaussian isotropic fields were investigated.
See also \cite{Dolgov_1999}.
Several results from this work can be summarized as follows:
\begin{itemize}
  \item The total number density of NPP is a constant depending only on the
        correlation radius of the signal.
  \item NPP can be classified into one of three kinds (knots, foci and seddles) depending on the
        local geometry of the field.
  \item The ratios of  number densities of these different kinds of NPP
        are fixed, independent of the power spectrum and depend only on the assumption
        of Gaussianity.
\end{itemize}
The third result motivates a simple test of Gaussianity, based on counting and classifying NPP and calculating the corresponding ratios of their counts.
In any finite sky area, random sampling error permits some departure of the actual counts from the theoretical number densities. Therefore, we compare the variation of the ratios across the sky with that of Gaussian simulations.

One additional feature is the properties of E- and B- modes of polarization, widely used in the search for cosmological gravitational waves.
In \cite{liu2018} it was shown that $Q$ and $U$ can be decomposed into $(Q_E, U_E)$ and $(Q_B, U_B)$ families, which generate the corresponding E and B-modes: $(Q_E, U_E)\rightarrow E$ and $(Q_B, U_B)\rightarrow B$.
In this transform, we remain in the domain of the Stokes parameters, and the statistics of NPP apply to each.

In Section \ref{sec:2}, we review the theory of NPP in the polarization field and derive the main analytical results.
In \ref{sec:3}, we summarize the procedure for measuring NPP in pixelized data. 
The theory of NPP in the E and B-modes is discussed in Section~\ref{sec:4}, with emphasis on the effect and detectability of lensing by this estimator.
In \ref{sec:5} the statistics of the singular point ratios are explored, and Gaussian simulations are compared to the Planck 2018 CMB maps \cite{planck2018a}.
A brief conclusion is given in \ref{sec:6}.

\section{Theory of non-polarized points}
\label{sec:2}

Inflation predicts Gaussian statistics for the CMB, including its polarization.
These assumptions allow us to also calculate predictions for the densities of non-polarized points in the CMB.
Unfortunately, the Planck CMB products (SMICA, Commander, NILC and SEVEM) are contaminated by foreground residuals and instrumental noise.
These factors will affect the statistics of NPP compared to a pure cosmological signal.
At the same time, these peculiarities of statistics can be used for determination of their morphology and amplitudes, and they can be masked out by more complicated filters designed in the domain of polarization intensity, e.g.~\cite{Liu:2019akp}.
We will discuss these anomalies in the next sections of the article, and now we turn to the theory for Gaussian signals, following \cite{naselsky1998}.
Note that this model is exact for statistically isotropic Gaussian noise added to the CMB. 

\subsection{Statistics of Gaussian Q and U and their derivatives}

We start with the basis for our description of the CMB polarization field from \cite{naselsky1998} for total $Q$ and $U$ without separation into E- and B-families.
We assume that $Q$ and $U$ are random Gaussian variables with possibly nonzero means and define the Stokes vector,
\begin{equation}
  \label{eqn:polvec}
  \vb*{P} = \mqty(Q \\ U), \quad \vb*{\mu} = \mqty(\mu_Q \\ \mu_U),
\end{equation}
where $\mu_Q$ and $\mu_U$ are the means of $Q$ and $U$. The covariance matrix of $Q$ and $U$ is
\begin{gather}
  \label{eqn:covmat}
  \vb*{\Sigma}_{QU} = 
  \begin{pmatrix}
    \sigma_{QQ} & \sigma_{QU} \\
    \sigma_{QU} & \sigma_{UU}
  \end{pmatrix}.
\end{gather}
Since only second-order correlations exist, the statistics of the Stokes parameters are described by a Gaussian distribution.
Using equations (\ref{eqn:polvec}) and (\ref{eqn:covmat}), we write the joint probability density for $Q$ and $U$:
\begin{equation}
  \label{eqn:QUdist}
  \mathcal{P}(Q, U) = \frac{1}{2\pi \det(\vb*{\Sigma}_{QU})^{\frac{1}{2}}}
  \exp[-\frac{1}{2} (\vb*{P} - \vb*{\mu})^T \vb*{\Sigma}_{QU}^{-1} (\vb*{P} - \vb*{\mu})].
\end{equation}

We now make our first assumptions that $Q$ and $U$ have the same variance, namely the spectral parameter $\sigma_0^2$.
We also take them to be completely uncorrelated, $\sigma_{QU} = 0$.
The mean of random variables with a Gaussian distribution does not affect the shape of the PDF, only its location in the space of random variables, so we take $\vb*{\mu} = 0$, and move to the polarization vector $\vb*{P} \to \vb*{P} + \vb*{\mu}$, such that $\vb*{P} - \vb*{\mu} \to \vb*{P} + \vb*{\mu} - \vb*{\mu} = \vb*{P}$. 
With these assumptions, equation~(\ref{eqn:QUdist}) simplifies to
\begin{equation}
  \label{eqn:QUdist-final}
  \mathcal{P}(Q, U) = \frac{1}{2\pi \sigma_0^2} \exp(-\frac{1}{2} \frac{I^2}{\sigma_0^2})
\end{equation}
where $I^2=||\vb*{P}||^2=Q^2+U^2$, and $\sigma_0^2$ is the variance of $Q$ and $U$.
For the purposes of analyzing NPP, we also want to describe the statistics of the field derivatives.
We define the Jacobian
\begin{equation}
  \label{eqn:the_jacobian}
  \vb*{J} = 
  \begin{pmatrix}
    Q_\theta & Q_\varphi \\
    U_\theta & U_\varphi
  \end{pmatrix}
\end{equation}
where
\begin{equation}
  Q_\theta = \frac{\partial Q}{\partial \theta}, \quad U_\theta = \frac{\partial U}{\partial \theta}, \quad Q_\varphi = \frac{\partial Q}{\partial \varphi}, \quad U_\varphi = \frac{\partial U}{\partial \varphi}.
\end{equation}
In constructing a probability density for the derivatives, similar assumptions are made as for the field components themselves.
Namely, that they are uncorrelated and have a variance tied to spectral parameters, $\sigma_{Q_iQ_i} = \sigma_{U_iU_i} = \sigma^2_1/2$.
All other variances are then zero.
Analogously to how a distribution was constructed for the $Q$ and $U$ components, we then write out the resulting probability density, by extending the standard covariance matrix with that of the field derivatives.
\begin{widetext}
\begin{equation}
  \label{eqn:joint_distribution}
  \mathcal{P}(Q, U, Q_\theta, Q_\varphi, U_\theta, U_\varphi) =
  \frac{4}{(2\pi)^3 \sigma_0^2 \sigma_1^4}
  \exp[
    -\frac{1}{2} \left(
      \frac{Q^2 + U^2}{\sigma_0^2} + 2 \frac{Q_\varphi^2 + U_\varphi^2 + Q_\theta^2 + U_\theta^2}{
        \sigma_1^2
      }
    \right)
  ]
\end{equation}
\end{widetext}

\begin{figure*}[!th]
  \centering
  \includegraphics{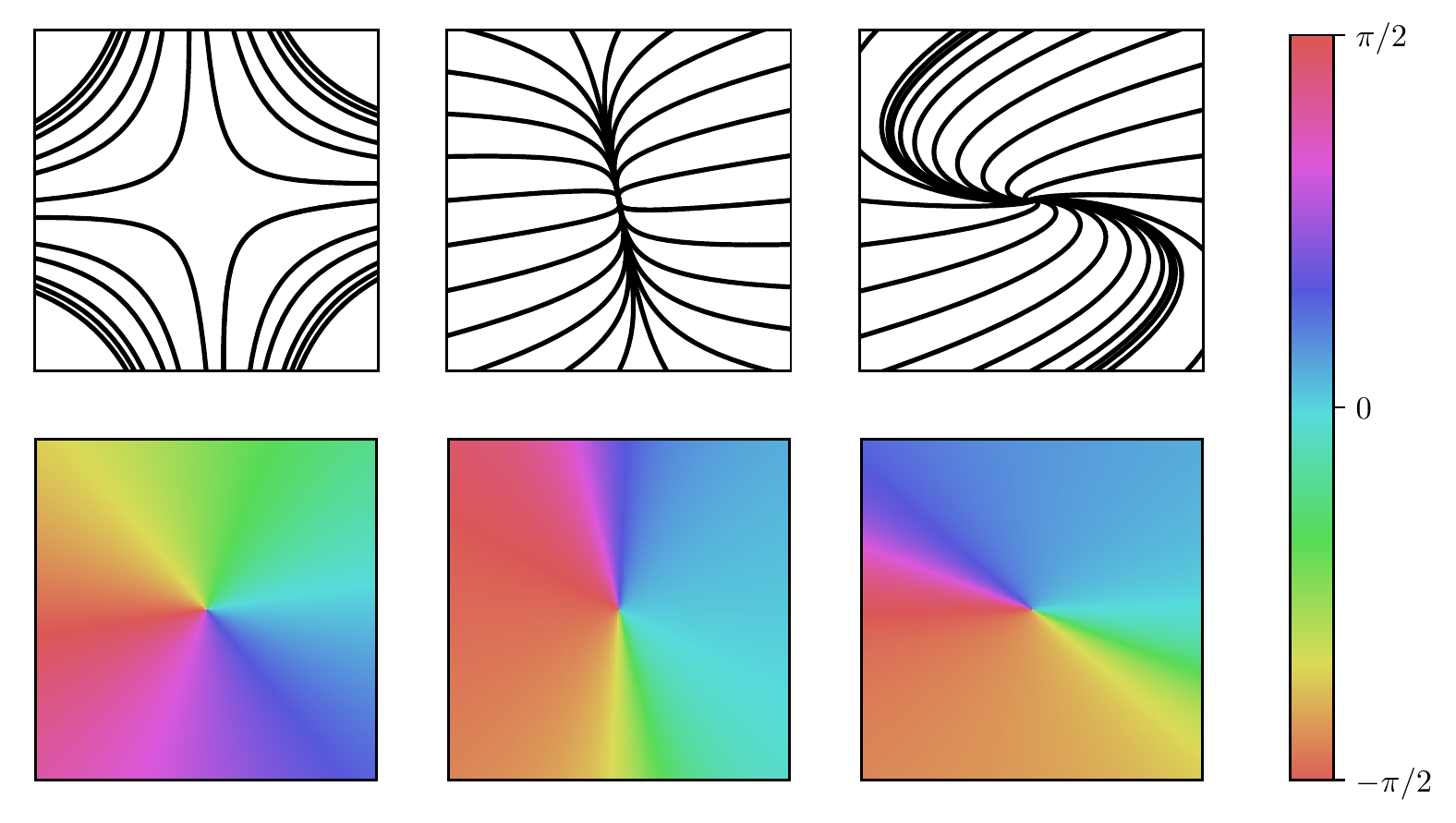}
  \caption{
    Samples of NPP for Gaussian $Q$ and $U$ in flat $(\theta, \varphi)$ space.
    The first column features a saddle, the second a knot and the last a focus.
    The second row is an image of the polarization angle $\Psi$ linearly interpolated around the singular point, based on the first-order derivatives of $Q$ and $U$ evaluated at the singular point itself and defining its classification.
  }
  \label{fig:sp_samples}
\end{figure*}

\subsection{Classification of non-polarized points}

Non-polarized points are defined to be places where the polarization vector from equation (\ref{eqn:polvec}) is zero, $\vb*{P} \equiv 0$.
This also means that at NPP, the polarization intensity, defined as $|\vb*{P}| = \sqrt{Q^2 + U^2}$, is zero.
Field dynamics around NPP are considered in the first order, based on the Jacobian of field derivatives.
A first-order change around the NPP is described by applying the Jacobian,
\begin{align}
  \vb*{P}(&\theta_0 + \dd\theta, \varphi_0 + \dd\varphi) =
  \vb*{P}(\theta_0, \varphi_0) + \dd\vb*{P} = \dd\vb*{P} \\ \nonumber &=
  \begin{pmatrix}
    \dd Q \\
    \dd U
  \end{pmatrix} = 
  \begin{pmatrix}
    Q_\theta & Q_\varphi \\
    U_\theta & U_\varphi
  \end{pmatrix} 
  \begin{pmatrix}
    \dd\theta \\
    \dd\varphi
  \end{pmatrix} = \vb*{J}
  \begin{pmatrix}
    \dd\theta \\
    \dd\varphi
  \end{pmatrix}.
\end{align}
The eigenvalues of the Jacobian end up completely determining the dynamics of the field and turning this into the usual exercise of the phase portraits around stationary points with the following characteristic equation for eigenvalues $\lambda$: 
\begin{align}
\det
 \begin{pmatrix}
    Q_\theta-\lambda & Q_\varphi \\
    U_\theta & U_\varphi-\lambda
  \end{pmatrix} =0
\label{lambda}
\end{align}
where
\begin{eqnarray}
\lambda_{1,2}=\frac{1}{2}\Tr(\vb*{J})\pm\sqrt{\left[ \frac{1}{2}\Tr(\vb*{J}) \right]^2-\det(\vb*{J})}.
\label{lambda1}
\end{eqnarray}

The trace and determinant of the Jacobian determine the classification of each singular point into either a focus, knot or saddle.
The conditions for foci, knots and saddles are respectively given by
\begin{gather}
\det(\vb*{J}) > \left[ \frac{1}{2}\Tr(\vb*{J}) \right]^2 > 0 \quad \text{(foci)}, \quad\nonumber\\
\left[ \frac{1}{2}\Tr(\vb*{J}) \right]^2 > \det(\vb*{J}) > 0 \quad \text{(knots)}, \nonumber \\ 
\det(\vb*{J}) < 0 \quad \text{(saddles)}.
\end{gather}
Trajectories of the polarization vector for each kind of NPP can be plotted around the singular point by solving the system of equations defined by the Jacobian in equation~(\ref{eqn:the_jacobian}) acting on a polarization state.
Examples are depicted in Fig.~\ref{fig:sp_samples}.

We next turn to discussion of the statistics of these NPP, based on \cite{naselsky1998}. 
The number density of different NPP is given by  \cite{naselsky1998}:
\begin{gather}
  N_f = \frac{\sqrt{2}}{16\pi r_c^2}
  \qquad
  N_k = \frac{1}{16\pi r_c^2} (2 - \sqrt{2})
  \qquad
  N_s = \frac{1}{8\pi r_c^2}
\end{gather}
These are the densities of foci, knots and saddles, respectively.
The radius of correlation is introduced, defined as $r_c = \sigma_0 / \sigma_1$.
Summing these densities gives the total singular point density:
\begin{gather}
  \label{eqn:total_density}
  N_{\text{total}} = N_f + N_k + N_s = \frac{1}{4\pi r_c^2}.
\end{gather}
Of note is that the ratios of these densities are parameter-independent.
We write them in terms of the focus density.
\begin{eqnarray}
  \frac{N_f}{N_k} = \sqrt{2} + 1
  \qquad
  \frac{N_f}{N_s} = \frac{\sqrt{2}}{2}
  \label{ratio}
\end{eqnarray}
Following the arguments here, a statistical ensemble of the  pure Gaussian CMB skies should be one in which these ratios hold, and measuring these ratios in data allows us to describe the statistics of a CMB.
However, for a single realization of the CMB sky with some finite number of NPP, the actual detected ratios are random variables. Thus, our goal is to extend the analysis of \cite{naselsky1998}
in order to determine the uncertainties of these ratios.

\begin{figure}[t]
    \centering
    \includegraphics{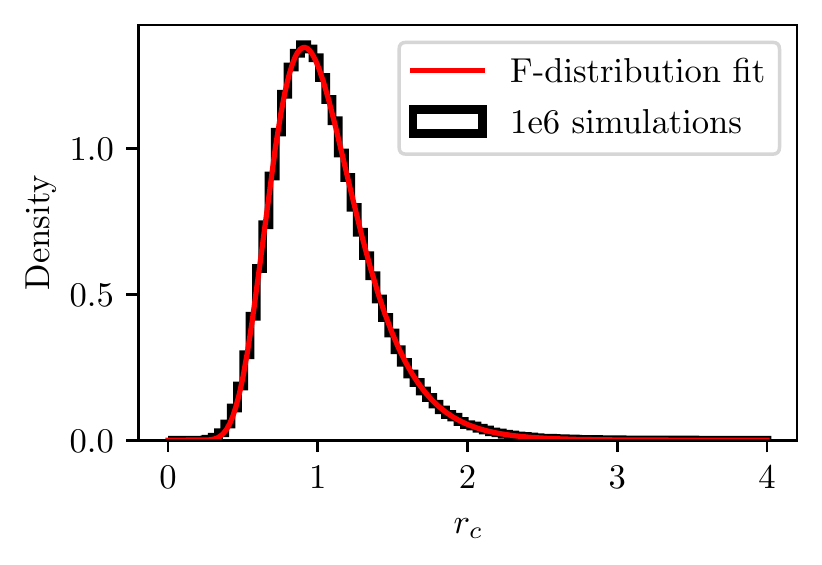}
    \caption{
        The correlation length $r_c$ follows an $F$-distribution. 
        Shown is the distribution of $r_c$ from $10^6$ Gaussian simulations compared to an $F$-distribution fit with parameters $\alpha = \beta \approx 42$. At high resolution, the $F$-distribution reduces to a Cauchy distribution.
    }
    \label{fig:fdist}
\end{figure}

Assuming that the underlying $Q$ and $U$ data are Gaussian random variables, we can derive the distribution followed by the ratios.
Let $\hat{\sigma}_0^2$ be the sample variance of $Q$ and $U$.
It follows a chi-squared distribution.
$\hat{\sigma}_1^2$ similarly is chi-squared distributed.
The correlation length, $r_c = \sigma_0 / \sigma_1$, is therefore the ratio of two chi-squared-distributed variables, which means it should follow (with suitable normalizations) an $F$-distribution.
We illustrate the distribution of the correlation length $r_c$ in Fig.~\ref{fig:fdist} for a statistical ensemble of $10^6$ Gaussian realizations and compare the result with the probability density function of an $F$-distribution, given by
\begin{eqnarray}
F(x,\alpha,\beta)=Ax^{\frac{\alpha}{2}-1}\left(1+\frac{\alpha}{\beta}x\right)^{-\frac{\alpha+\beta}{2}}
\label{eq:f}
\end{eqnarray}
where $A$ is a normalization constant and $\alpha$ and $\beta$ are parameters of the distribution.

The analysis of NPP for Gaussian $Q$ and $U$ needs to be completed by description of the properties of polarization angle in the vicinity of these points. 
The behavior of $\Psi$, presented in the lower row of Fig.~\ref{fig:sp_samples}, is driven by the following equation:
\begin{equation}
    \frac{d}{d\Theta} \tan(2 \Psi) = \frac{\det(\vb*{J})}{Q^2},
\end{equation}
where $\Theta$ is the polar angle in the coordinate system centered on the non-polarized point.
From this equation we can explain part of the behaviour observed in Fig.~\ref{fig:sp_samples}: the sign of the determinant controls whether the polarization angle increases or decreases clockwise around the NPP.
For saddles, $\Psi$ decreases clockwise; for knots and foci, $\Psi$ increases.

\section{Detecting non-polarized points in intensity maps}
\label{sec:3}

\begin{figure*}[t]
  \centering
  \includegraphics[width=0.98\textwidth]{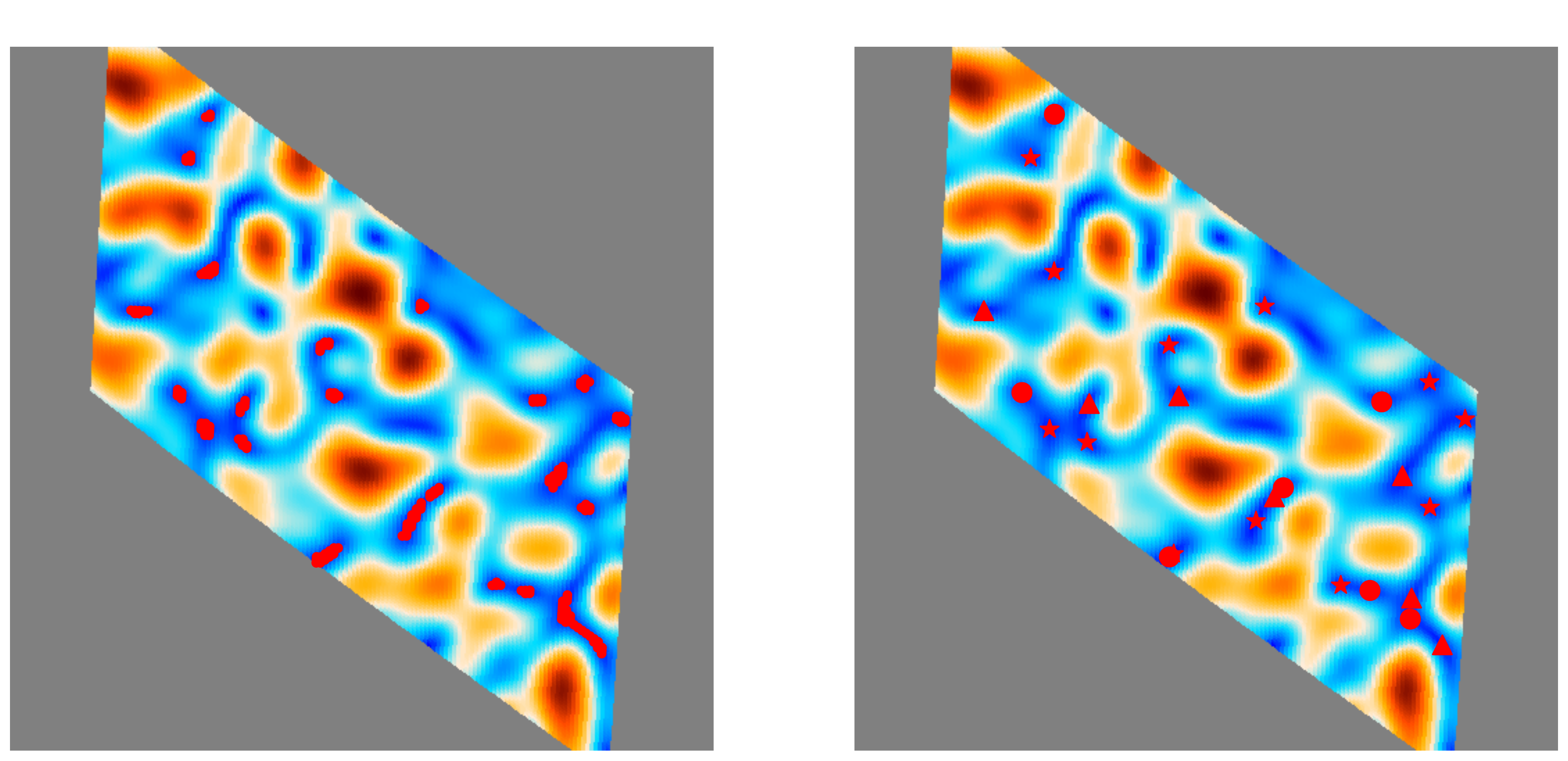}
  \caption{
    Finding and classifying NPP in a sample mother pixel with $N_\mathrm{side} = 16$ from Gaussian simulations.
    In the left panel are highlighted sufficiently cold areas in the polarization intensity
    that also contain crossing points in both $Q$ and $U$. In the right
    panel are the NPP that are kept for analysis of ratios.
    Circles are foci, triangles are knots and stars are saddles.
    In the right panel there are 7 foci, 7 knots and 12 saddles.
    The ratios of this subarea are $N_f / N_k = 1$ and $N_f / N_s = 7/12$.
  }
  \label{fig:sps_sample_subarea}
\end{figure*}

Non-polarized points were defined as the places where both $Q$ and $U$ are zero.
For the practical analysis of NPP we need to implement some filtration for the Planck data in order to restore approximate continuity of the discrete pixelized data.
We will use a standard Gaussian filter with a FWHM of $0.5$ degrees.
Since these data are very noisy, to some extent the Gaussian filter will remove high multipoles from the analysis and effectively decrease the power of the noise.
Solving for NPP in the $Q$, $U$ field configuration would be complicated, as the extremal points can be minima, maxima, or saddles.
So we instead look to the polarization intensity, that is bounded from below by 0, where the singular points should be.
Floating point inaccuracies and the resolution of available data further complicate the search, the true NPP will not appear as perfect zeroes in the data.
As such, any local minimum in the polarization intensity with a sufficiently small value could be a candidate for a singular point. 
To deal with this, we introduce the parameter $\varepsilon$ with the same units as the polarization intensity.
This parameter is chosen according to the correlation scale of the map.
We then select contiguous sub-threshold basins, within which both $Q$ and $U$ separately change sign, as candidates for NPP.
Local minima pixels within each basin are determined using a recursive search, and duplicate minima that are nearly adjacent are filtered out.
The remainder are considered NPP and used for calculating ratios inside the region in question.
Therefore the conditions for a NPP at a particular pixel $\vb*{n}_0$ are:
\begin{enumerate}
    \item The polarization intensity satisfies $I(\vb*{n}_0) < \varepsilon$ for an appropriately chosen value of $\varepsilon$.
    \item The polarization intensity is a local minimum, meaning $I(\vb*{n}_0) < I(\vb*{n}_i)$ for all pixels $\vb*{n}_i$ adjacent to $\vb*{n}_0$.
    \item Both positive and negative values of $Q$ and $U$ are found in the connected sub-$\varepsilon$ region including $\vb*{n}_0$.
    \item There are no other candidates also meeting conditions 1--3 with $I < I(\vb*{n}_0)$ within an angular distance $\alpha_\mathrm{max}$ of $\vb*{n}_0$.
\end{enumerate}
$\varepsilon$ and $\alpha_\mathrm{max}$ are parameters of the search method. 
Once the pixels corresponding to each NPP are determined, they can be classified by simply evaluating the Jacobian at each pixel and applying the rules described above in Section \ref{sec:2}.
The code that performs this detection of singular points is available at \cite{libpol}, and an illustrative example of NPP detection and classification is shown in Fig.~\ref{fig:sps_sample_subarea}.

It is convenient to count NPP in a target sky that has been divided into subareas.
In any finite sky area, the actual ratios of counts will deviate from the Gaussian prediction.
By dividing the sky into different areas and calculating the ratio within each, we can effectively run many "simulations" in one sky, and compare the resulting variation to that of Gaussian simulations.
This results in an estimator which is a more informative than calculating the full-sky ratio alone.

For data serialized in the HEALPix \cite{healpix} format, a HEALPix pixel at some fixed NSIDE is the natural choice.
What remains is the choice of $\NSIDE$. For statistical purposes, it should be large enough that a distribution of evaluated ratios from each of the subareas is available for analysis, yet not too large that clustering algorithms take up a lot of memory and CPU time.

In this work we settled on choosing $\NSIDE=8$ and $16$ for the mother pixel, preserving the pixelization for the sky map $N_\mathrm{side}=2048$.
This choice of overpixelization determines the ``mother pixels'' within each of which NPP are counted and ratios calculated to derive the final distribution functions.
In practice, however, the detection algorithm is run on the full sky maps, and afterwards the points are assigned to their mother pixels.
This means that the algorithm is not at risk of edge effects.

\section{E/B-family decomposition and the total number density of non-polarized points}
\label{sec:4}

As we have pointed out above, the decomposition of the Stokes parameters into E- and B-families provides a direct link to the corresponding E- and B-modes of polarization (see for details \cite{liu2018}).
Since all these transforms are based on liner combination of $Q$ and $U$ parameters, both E- and B-families will preserve the statistical properties of $Q$ and $U$.
Thus, Gaussian $Q$ and $U$ will generate Gaussian $Q_E,U_E$ and $Q_B,U_B$, and all the criteria for NPP, presented in Section~\ref{sec:3}, will be fully applicable for the E- and B-families.  

\begin{figure*}[t]
    \centering
    \includegraphics{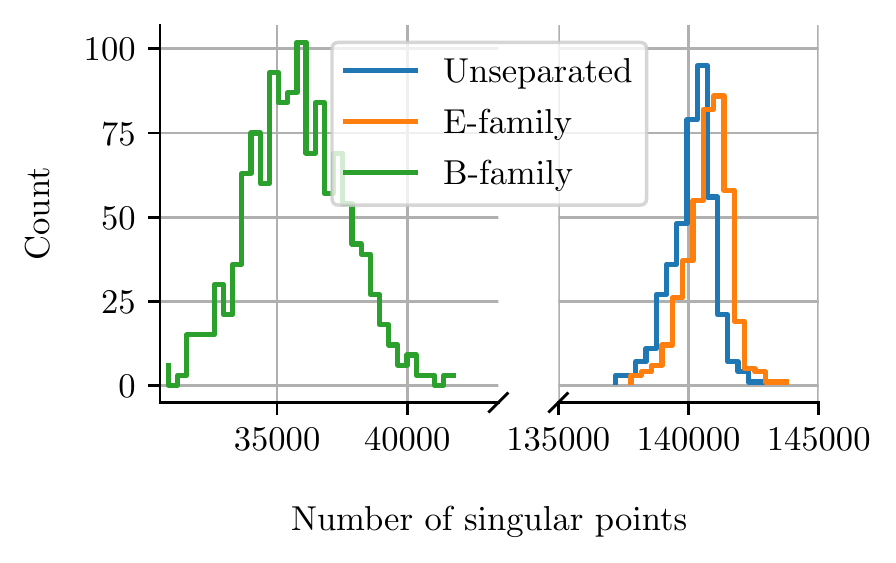}
    \includegraphics{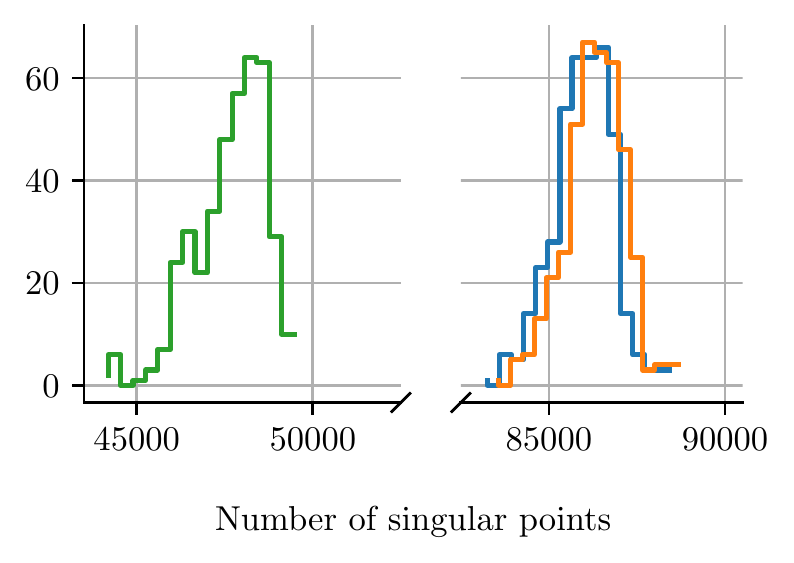}
    \caption{Range of the number of non-polarized points in simulations with $r = 0.05$ and 0.5-degree smoothing (see Section IV for details). The blue lines are for the full unseparated polarization map, and the orange lines are for the E-family only. The B-family is shown with green lines in the left-offset panels. The left pair of panels are based on simulations without lensing, and the right pair of panels include lensing. Simulations for these results were created in CAMB using parameters derived from the Plik likelihood \cite{planck2020}.}
    \label{EBN}
\end{figure*}

We begin with the number density of NPP. 
In Fig.~\ref{EBN} we show the number of counts for NPP for E- and B-families, taken from random realisations of best fit Planck 2018 $\Lambda$CDM model with tensor to scalar ration $r=0.05$ and with lensing of E-mode included.
We used smoothing of the simulated maps with Gaussian kernel with FWHM $0.5$ degrees.

As it is seen from Fig.~\ref{EBN}, the distribution of total number of singular points for unseparated signal and the corresponding E/B-families are approximately Gaussian with a small separation of the means for unseparated and E-family, and a strong difference between the E- and B-families.
Note a very interesting effect, that the point of maximum of NPP for the E-family is slightly bigger than for the unseparated signal. 
This phenomenon can be understood in terms of the correlation radius $r_c$ from Eq.~\ref{eqn:total_density}.

For unseparated signal $(Q,U)$, the correlation radius is given by the ratio $r_c=\sigma_0/\sigma_1$, where 
\begin{align}
    \Delta^2&=2\sigma_0^2=\expval{(Q_E+Q_B)^2+(U_E+U_B)^2}=\Delta^2_E+\Delta^2_B,\\
    \sigma^2_1&=\gamma^2_E+\gamma^2_B,
\label{EBN1}
\end{align}
and $\Delta^2_E=\expval{Q^2_E}+\expval{U^2_E}$ and $\Delta^2_B=\expval{Q^2_B}+\expval{U^2_B}$ are the corresponding variances for amplitudes of E/B-families, and $\gamma^2_E, \gamma^2_B$ are the variances for the first derivatives:
 $\gamma_E^2 = 4(\expval{(Q_E')^2} + \expval{(U_E')^2})$ and $\gamma_B^2 = 4(\expval{(Q_B')^2} + \expval{(U_B')^2)}$.
So, for the whole sky the total number of unseparated NPP can be represented as follows:
\begin{align}
N_\mathrm{tot} &=\frac{\gamma^2_E+\gamma^2_B}{\Delta^2_E+\Delta^2_B}\simeq N_\mathrm{tot}^E\left(1+\frac{\gamma^2_B}{\gamma^2_E}-\frac{\Delta^2_B}{\Delta^2_E}\right)\nonumber\\
    &=N_\mathrm{tot}^E\left[1-\frac{\Delta^2_B}{\Delta^2_E}\qty(1-\frac{N_\mathrm{tot}^B}{N_\mathrm{tot}^E})\right].
\label{EBN2}
\end{align}
Thus, for $\frac{N_\mathrm{tot}^B}{N_\mathrm{tot}^E}<1$ the total number of unseparated NPP is lesser than for E-family, which one can see reflected in the Gaussian simulations presented in Fig.~\ref{EBN}.
Also, from equation~(\ref{EBN2}) we can see that the size of the deviation will grow according to $\Delta_B^2$, and therefore also according to $r$, the tensor-to-scalar ratio.

At the end of this section we would like to discuss the dependence of the total number of NPP for the B-family on the tensor/scalar ratio $r$, based on analysis of pure Gaussian simulations.
We will exploit the fact that for the primordial CMB signal without lensing of the E-mode, the total number of NPP $N_\mathrm{tot}^B$  depends only on the ratio $ \gamma^2_B/\Delta^2_B$, where both numerator and denominator are proportional to $r$. Thus, $N_\mathrm{tot}^B$ should not depend on $r$ without incorporation of the lensing effect.

From a theoretical point of view, we may expect that with weak lensing of the E-mode, the number of NPP will be practically the same as without lensing for all realistic values of $r$.
For the B-family, the dependency $N_\mathrm{tot}^B$ on $r$ is non-trivial due to the following reasons.

Firstly, it should critically depend on the balance between theoretical (without smoothing by the antenna beam or any others filters) correlation radius $r_\mathrm{cmb}$ and the effective scale of smoothing $\Theta_\mathrm{sm}$.
If $\Theta_\mathrm{sm} \ge r_\mathrm{cmb}$, then the $\gamma^2_B$-parameter is still proportional to $r$ for the CMB B-family and it is affected by the contribution of the lensed E-component:
\begin{equation}
\gamma^2_B\simeq \gamma^2_{B,\mathrm{cmb}}(r)+\gamma^2_{B,\mathrm{lens}},
\label{lens}
\end{equation} 
where the first term corresponds to the cosmological B-mode with tensor to scalar ratio $r$, and the second one is for the lensing effect (almost independent of $r$).
The same representation is valid for the variances:
\begin{equation}
\Delta^2_B\simeq \Delta^2_{B,\mathrm{cmb}}(r)+\Delta^2_{B,\mathrm{lens}}.
\label{lens1}
\end{equation}
Taking into account that $\gamma^2_{B,\mathrm{cmb}}(r)$ and $\Delta^2_{B,\mathrm{cmb}}(r)$ both are proportional to $r$, we can represent these terms in the following way:
\begin{equation}
\Delta^2_{B,\mathrm{cmb}}(r)=\frac{r}{r_*}C,\hspace{0.5cm}\gamma^2_{B,cmb}(r)=\frac{r}{r_*}D
\label{lens2}
\end{equation}
where $r_*$ is some arbitrary normalisation parameter, and $C$ and $D$ correspond to $\Delta^2_{B,\mathrm{cmb}}(r)|_{r=r_*}$ and $\gamma^2_{B,\mathrm{cmb}}(r)|_{r=r_*}$.

Secondly, the total number of NPP $N_\mathrm{tot}^B$ critically depends on the ratios between the CMB and the lensing terms in equations~(\ref{lens}--\ref{lens1}).
Namely, 
\begin{equation}
N_\mathrm{tot}^B(r)=\frac{\frac{r}{r_*}D+\gamma^2_{B,\mathrm{lens}}}{\frac{r}{r_*}C+\Delta^2_{B,\mathrm{lens}}}
\label{lens3}
\end{equation}
whence 
\begin{equation}
N_\mathrm{tot}^B(r) \xrightarrow[r \to 0]{} \frac{\gamma^2_{B,\mathrm{lens}}}{\Delta^2_{B,lens}}=N_{B,\mathrm{lens}}
\end{equation}
in the small-$r$ limit, and 
\begin{equation}
N_\mathrm{tot}^B(r) \xrightarrow[r \to \infty]{} \frac{D}{C} = N_{B,\mathrm{cmb}}
\end{equation}
when $r$ is large.
Thus, the entire range of variation of $N_\mathrm{tot}^B(r)$ is reduced to a transition between two constants.
The details of this transition can be traced as follows.
Let's assume that the CMB tail of equation~(\ref{lens3}) dominates over the lensed part: $\frac{r}{r_*}D\gg \gamma^2_{B,\mathrm{lens}}$ and $\frac{r}{r_*}C\gg \Delta^2_{B,\mathrm{lens}}$.
In this case the total number of NPP is given by the following asymptotic:
\begin{equation}
\frac{N_\mathrm{tot}^B(r)}{N_{B,\mathrm{cmb}}}\simeq1-\frac{\Delta^2_{B,\mathrm{lens}}}{C}\left(\frac{r_*}{r}\right)\left(1-\frac{N_{B,\mathrm{lens}}}{N_{B,\mathrm{cmb}}}\right)
\label{lens4}
\end{equation}
Here $N_{B,\mathrm{lens}}/N_{B,\mathrm{cmb}}\ll 1$ and the total number of NPP is slightly smaller than for pure CMB.

From a practical point of view, of greatest interest is the asymptotics $N_{B,\mathrm{lens}}/N_{B,\mathrm{cmb}}\gg 1$, when the structure of the polarized B-signal is completely determined by the lensing effect of E-mode.
This effect will dominate at $r \ll 0.01$, which is a target for the forthcoming CMB experiments. 

\begin{figure}[t]
  \centering
    \includegraphics{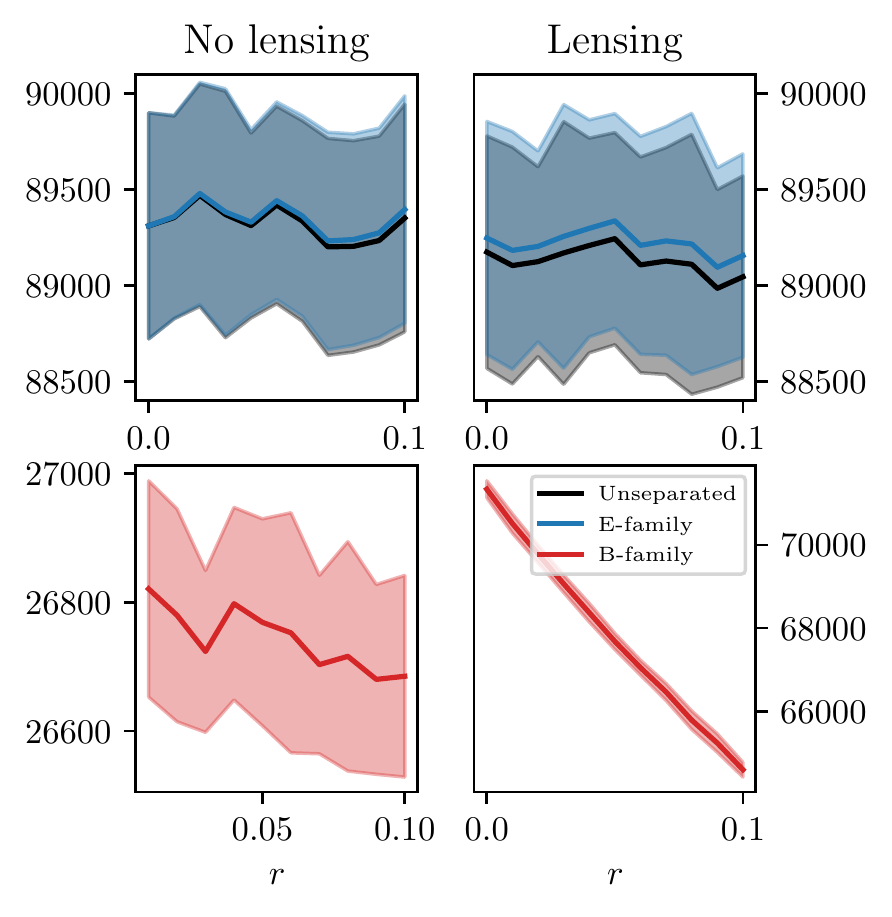}
  \caption{The results of numerical simulation for total numbers of NPP for unseparated signal (gray), E (blue) and B-families (red) 
  with and without lensing. The left column corresponds to the unlensed case and the right column is for lensed.
  The width of the contours corresponds to 68\% confidence level. The solid lines correspond to the mean over realisations.
  }
  \label{fig:r}
\end{figure}

For $N_{B,\mathrm{lens}}/N_{B,\mathrm{cmb}}\gg 1$ we get:
\begin{eqnarray}
\frac{N_\mathrm{tot}^B(r)}{ N_{B,\mathrm{lens}}}\simeq 1-\left(\frac{r}{r_*}\right)\frac{C}{\Delta^2_{B,\mathrm{lens}}}
\left(1-\frac{N_{B,\mathrm{cmb}}}{N_{B,\mathrm{lens}}}\right)
\label{lens5}
\end{eqnarray}
Confirmation of the asymptotics in equation~(\ref{lens5}) can be seen in Fig.~\ref{fig:r}, where we give results of numerical simulations for total numbers of NPP for the unseparated signal, E- and B-families  with and without lensing.
To generate an ensemble of Gaussian realizations, we used the following parameters from Planck 2018 date release in standard notation \cite{planckparameters2020}: $H_0=67.5$ km/sMpc, $\Omega_bh^2=0.022$, $\Omega_ch^2=0.122$, $A_s=210^{-9}$, $\tau=0.06$ and $n_s=0.965$.

\section{Non-polarized points in the 2018 Planck maps}
\label{sec:5}

\subsection{Contamination of the 2018 Planck maps}

Both the theory and computational methods for counting non-polarized points can now be
applied to the latest Planck release. We also get a baseline for these methods by
comparison with many Gaussian realizations from a best-fit power spectrum with
$r=0.05$. With the method of NPP we can measure how well each Planck
map matches the Gaussian expectation. The Kullback-Leibler entropy \cite{kullback1951} was also
used as an indicator for how well the Planck maps matched the Gaussian realizations.

\begin{figure*}[t]
\includegraphics{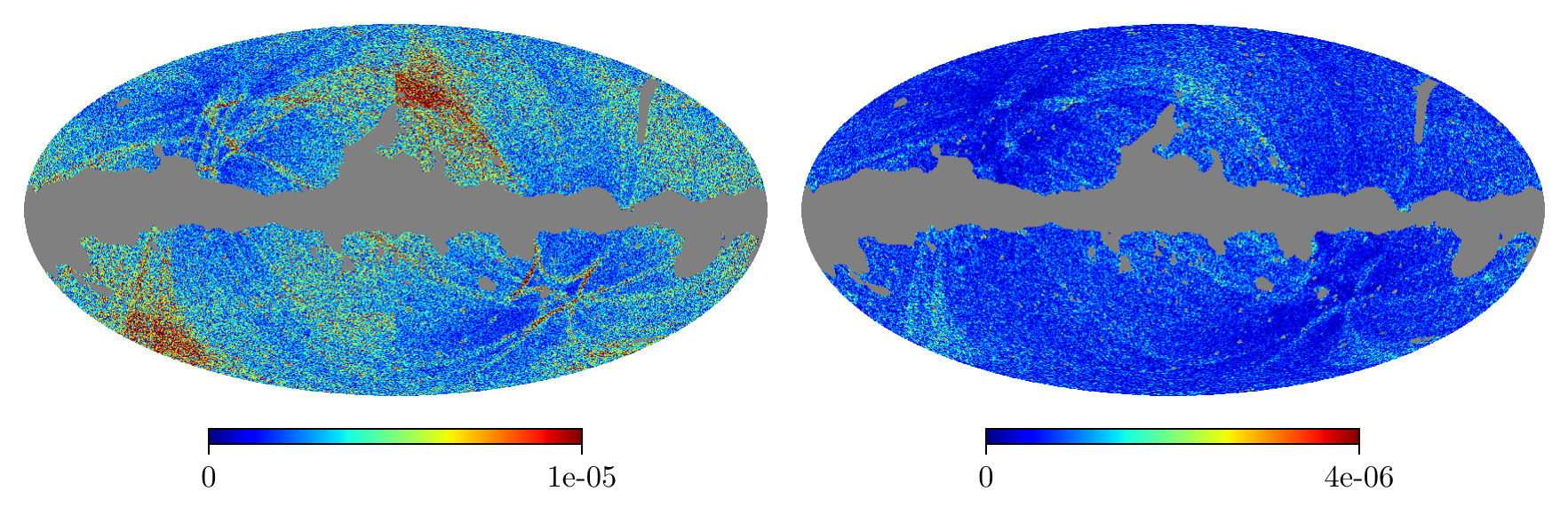}
 \caption{The maps of differences between SMICA and Commander, i.e.\ $\sqrt{(Q_\mathrm{SMICA} - Q_\mathrm{Comm})^2 + (U_\mathrm{SMICA} - U_\mathrm{Comm})^2}$, for $N_\mathrm{side}=2048$ (left panel) and $N_\mathrm{side}=128$ (right panel). }
   \label{diff1}
\end{figure*}

An important feature of Planck's polarization data is a relatively small signal-to-noise ratio (SNR) for the E-mode (SNR $\sim 1$), and absence of detection for the primordial  B-mode.
In addition, when analyzing the Planck data, one cannot ignore the effects of systematics, which can interfere with the instrumental noise and the foreground residuals. 
For illustration of the morphology of these non-cosmological components of the derived Planck 2018 CMB products in Fig. \ref{diff1} we show the intensity of the difference between SMICA and
Commander maps, defined as 
\begin{equation}
\Delta I=\sqrt{(Q_\mathrm{SMICA} - Q_\mathrm{Comm})^2 + (U_\mathrm{SMICA} - U_\mathrm{Comm})^2}
\end{equation}

Both these maps were taken with maximal resolution $N_\mathrm{side}=2048$. In the map of difference  SMICA--Commander, the cosmological
signal is removed and  this map clearly illustrates the interference between the instrumental noise and scan strategy patterns. Note that reduction of angular resolution down to $N_\mathrm{side}=128$ decreases the
amplitude of the noise, but preserves the morphology of contaminants. 

Thus, in reality, applying NPP analysis to the Planck CMB products, we are not directly testing Gaussianity of the primordial
CMB alone, but also potential contamination of the component separation
procedure. As a simple model, we can write the measured polarization field as
\begin{align}
 Q_\mathrm{pl} &= Q_\mathrm{cmb} + Q_R + Q_\mathrm{noise}=Q_\mathrm{cmb}+Q_N,\nonumber\\
 U_\mathrm{pl} &= U_\mathrm{cmb} + U_R + U_\mathrm{noise}=U_\mathrm{cmb}+U_N,
\label{noise}
\end{align}
where $Q_\mathrm{cmb},U_\mathrm{cmb}$ is the primordial signal, $Q_R,U_R$ are the residuals of systematics and
foregrounds, and $Q_\mathrm{noise},U_\mathrm{noise}$ are  noise. The NPP we measure are those where
$Q_\mathrm{pl} = 0, U_\mathrm{pl}=0$, i.e. $Q_\mathrm{cmb}= -Q_N, U_\mathrm{cmb}=-U_N$.  Any nonzero residuals or noise will
therefore perturb the location of the NPP compared to the
primordial signal. Also, note that if the residuals are correlated with the
CMB component or with the noise component, this can influence the location and density of NPP in the Planck maps.
As we have pointed out in Section II, the effect of existence of NPP is not local (see Fig.~\ref{fig:sp_samples}). In the vicinity of each CMB NPP the morphology of the signal can be
represented as follows:
\begin{align}
Q_\mathrm{pl} &\simeq Q'_xx +Q'_yy +Q_N(x,y),\nonumber\\
U_\mathrm{pl} &\simeq U'_xx +U'_yy +U_N(x,y)
\label{noise1}
\end{align}
where $x$ and $y$ correspond to the $\theta$ and $\phi$ coordinates around NPP, centered at $x=0,y=0$, and $Q'_x,Q'_y,U'_x,U'_y $ are the corresponding derivatives at that point. 

At the point $x=0,y=0$ the CMB component vanishes, while $Q_N(0,0)\neq 0,U_N(0,0)\neq 0 $.
Thus, at CMB NPP the structure of the polarization  for Planck noisy maps will be destroyed by the noise. However, linear behaviour of the CMB signal still will be preserved in the area
around NPP, when the terms linear in $x,y$ will dominate over the noise component. Qualitatively, the
corresponding scale can be estimated as follows. The linear terms for $Q_\mathrm{pl},U_\mathrm{pl}$ in equation~(\ref{noise1}) are of the order of 
\begin{equation}
A\sim \sigma_1\eta\simeq \sigma_0\frac{\eta}{r_c}
\label{noise2}
\end{equation}
where $\eta$ stands for the $x$ or $y$ coordinates. The noise term in equation~(\ref{noise1}) is about $\sigma_N$, where $\sigma_N$ is RMS of noise. Thus $A>B$ leads to the following constraint on $\eta$:
\begin{eqnarray}
\eta\ge r_c\frac{\sigma_N}{\sigma_0}\sim r_c (SNR)^{-1}
\label{noise3}
\end{eqnarray}
Thus, if signal-to noise ratio $\text{SNR} \gg 1$, the structure of the signal around CMB NPP is very well detectable for 
\begin{eqnarray}
 r_c (SNR)^{-1}\le \eta\le n r_c
\label{noise4}
\end{eqnarray}
where $n$ is $1$ to $3$, and marginally, it can be seen even for $\text{SNR} \sim 1$ to $2$.

\begin{figure*}[!ht]
  \centering
  \includegraphics{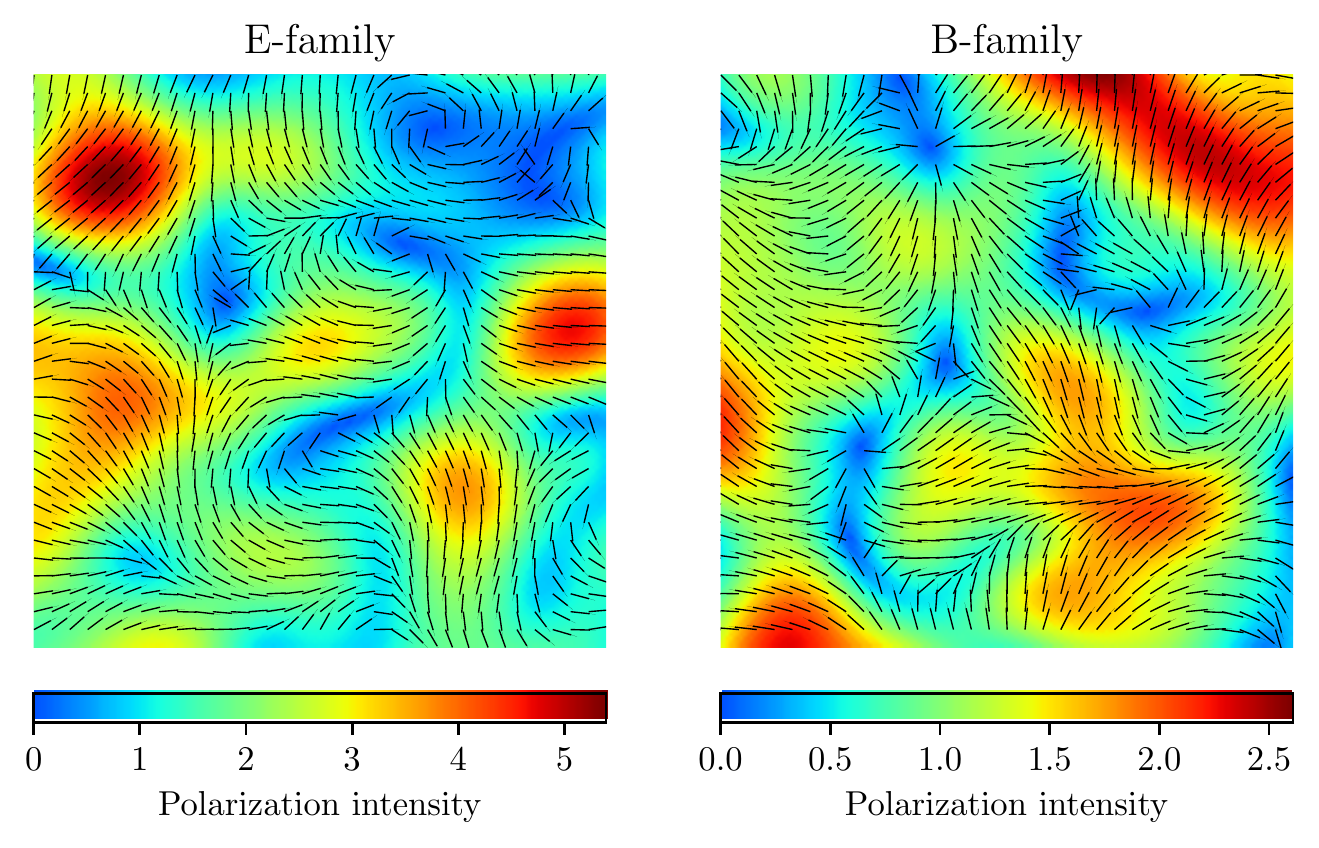}
  \includegraphics{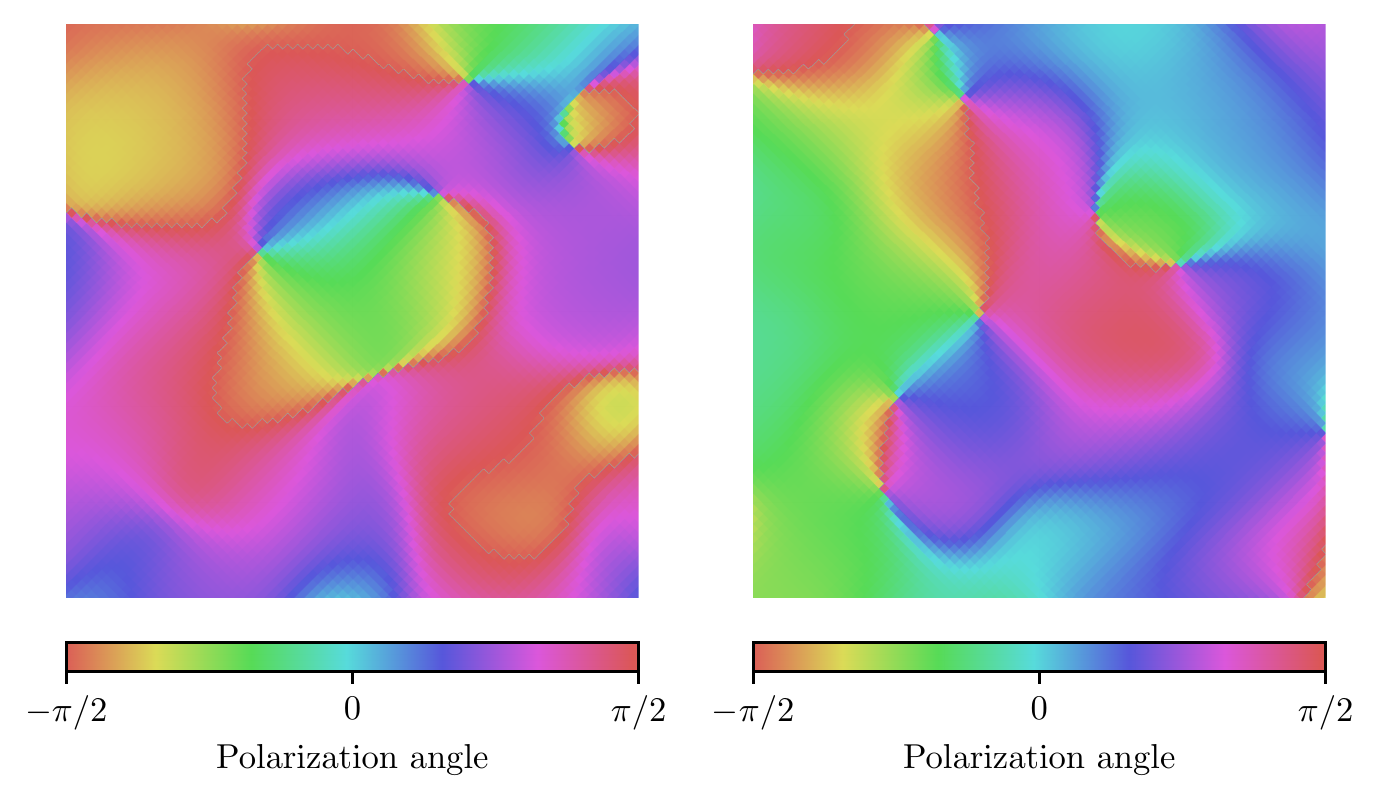}
  \caption{Top row: E- and B-families of the Commander polarization maps for $Q$ and $U$. The background of the maps corresponds to the intensity for each family (in $\mu K$). Orientation of the Stokes vector is indicated by black lines. The Commander map has angular resolution $N_\mathrm{side}=2048$ with Gaussian smoothing by $0.5^\circ$. The bottom row shows the polarization angle for the same maps correspondingly. The non-polarized points are clearly visible in this representation.}
  \label{decomp}
\end{figure*}

In Fig.~\ref{decomp} we show some $2.5^\circ \times 2.5^\circ$ patch of the Planck 2018
Commander polarization map, decomposed into E- and B-families and smoothed by a Gaussian filter with FWHM of 0.5 degrees. 
The maps of intensities for the E- and B-families indicate the position of the NPP as points of minima, while the 
line segments show the polarization angles. The two bottom maps illustrate
the morphology of the maps of polarization angles of the E- and B-families in detail. From the maps
of intensity (the top row) one can find that all zones colored by deep blue correspond to the 
NPP. The bottom row maps show the anomalies of the polarization angle in the vicinity of these
points in full agreement with theoretical expectations (for comparison see Fig.~\ref{fig:sp_samples}). Taking into account that $2.5^\circ \times 2.5^\circ$ maps in Fig.~\ref{decomp} correspond to the fraction of the sky $f_\mathrm{sky}\simeq 1.5 \times 10^{-4}$, from Fig.~\ref{EBN} one can find that the number of NPP in Fig.~\ref{decomp} is in agreement with
theoretical expectations as well.

\subsection{Ratio distributions for the Planck maps}

In this section we will address the problem of ratios for knots, foci and saddles 
(see equation~(\ref{ratio})) for the Planck 2018 SMICA, Commander, NILC and SEVEM maps. An important
feature of this test is that it depends only on statistical properties of the signal (CMB plus noise) and if noise is the dominating part of the signal, it will reveal important information about itself. For evaluation of statistical significance of the ratio test 
it is possible to use any simulations: for example, Planck FFP 10 or pure Gaussian.

\begin{figure*}[t]
  \centering
  \includegraphics[width=0.98\textwidth]{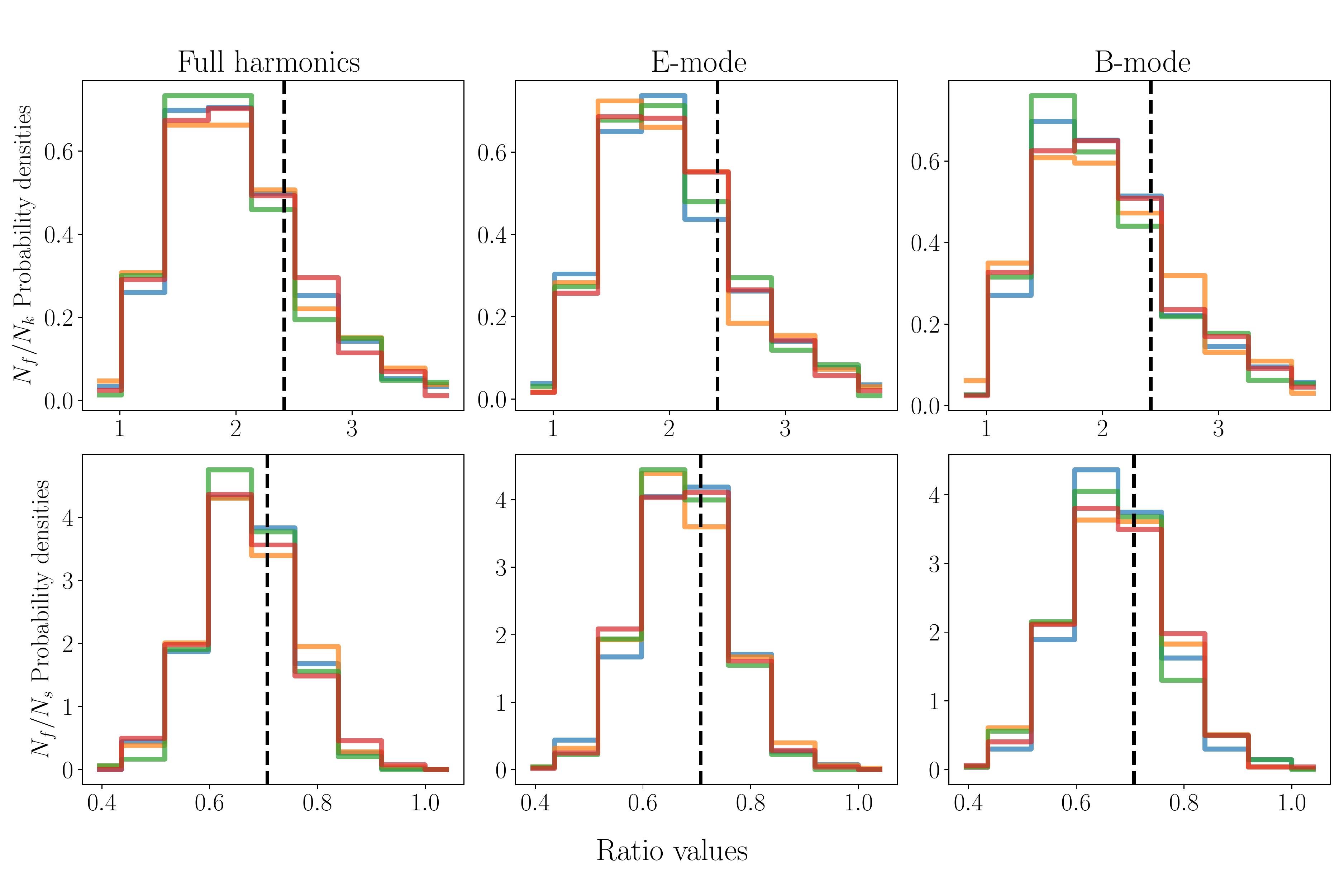}
  \caption{
    Ratio distributions for skies decomposed into separate sets of E/B-families. From left to right each column corresponds to full harmonics, the E-mode and
    the B-mode. The top row corresponds to the $f/k$ ratio, and the bottom to the $f/s$
    ratio. Each color corresponds to one of the Planck maps. Blue is SMICA, yellow is NILC, green
    is SEVEM and red is Commander.
  }
  \label{fig:eb_decomp}
\end{figure*}

For the ratio test all Planck maps and Gaussian simulations of the sky were smoothed by
Gaussian filter with FWHM 0.5 degrees and 
 divided into subareas (mother pixels)  of HEALPix pixels at $\NSIDE$ 8 and 16.
In each subarea outside the common mask, all NPP were counted and ratios were evaluated. These
were used to construct density histograms (see Fig.~\ref{fig:eb_decomp}) with the aim of comparing them to Gaussian
simulations. An important choice in creating histograms is a choice of the number
of bins.

The simple Sturges \cite{Sturges} estimator of $\log_2(n) + 1$ was used 
to provide a suitable number of bins for analyzing the data, where $n$ is the number
of data points available. The Doane estimator \cite{Doane} was also tried to compensate
for possible skewing in the data, but it gave the same results as Sturges, so the
latter was used. Each subarea contributes one data point for analysis, and
given the NSIDEs we were working at, we had at most either 768 ratio values for $\NSIDE=8$
or 3072 ratio values for $\NSIDE=16$. Based on this, the Sturges estimator provides 9 to 11
as the range for suitable numbers of bins. Bin edges were chosen accordingly to
encompass every realization of the sky under analysis. A cutoff value for the ratio value
was used to ignore statistical outliers, subareas with an unlikely small number of NPP.

Constructing these ratio distributions show that they are close to the theoretical expectations for a Gaussian random process.
Precise properties of each distribution
are given in Table~\ref{tab:ratio_stats}.

\begin{table}[t]
  \centering
  \begin{tabular}{ c|c|c|c|c }
    \hline 
    Map name & ex$(f/k)$ & stdev$(f/k)$ & ex$(f/s)$ & stdev$(f/s)$ \\ 
    \hline 
    Gaussian & 2.0349 & 0.3240 & 0.6763 & 0.0071 \\ 
    SMICA & 2.0149 & 0.3357 & 0.6716 & 0.0076 \\ 
    NILC & 2.0143 & 0.3756 & 0.6719 & 0.0084 \\ 
    SEVEM & 1.9934 & 0.3377 & 0.6702 & 0.0067 \\ 
    CMDR & 2.0073 & 0.3190 & 0.6712 & 0.0087 \\
  \end{tabular}
  
  \caption{
    Statistical information for all Planck maps and a ratio distribution from many Gaussian simulations.
    All data are calculated from the density histograms in Fig.~\ref{fig:eb_decomp}, at the same fidelity
    of 9 bins. Given are the expectation value (ex) and standard deviation (stdev).
    The theoretical ratio values to as many significant digits are 2.4142 and 0.7071 for the
    $f/k$ and $f/s$ ratios, respectively.
  }
  \label{tab:ratio_stats}
\end{table}

\subsection{Comparison with Gaussian distributions}

As an empirical test for the Planck maps, the Kullback-Leibler entropy between each of the Planck maps and a
background Gaussian was calculated. The KL entropy is a measure of the divergence between a sample
probability distribution $P$ and a background distribution $Q$. For discrete distributions,
it is defined as
\begin{gather}
  \Theta_{KL}(P|Q) = \sum_i P_i \ln(\frac{P_i}{Q_i}).
\end{gather}
Hence, the closer each $P_i$ is to each $Q_i$, the smaller the value of $\Theta_{KL}(P|Q)$ will
be and the better the match between the two distributions.

The KL entropy is a logarithmic estimator based on exponential probability densities that
have positive values in the whole domain. Due to this, discrete bins that end up having no data points
for some realizations of the sky contribute infinities to the entropy sum. To deal with this, we keep
track of the set of bins where a zero has occurred, and exclude those when calculating the KL-entropy.
Other methods like minimally smoothing the distributions and doping each bin with an extra data point
were considered, but they all performed the same. As such, dropping the zero bins was the preferred
method.

\begin{figure*}[t]
  \centering
  \includegraphics[width=0.88\textwidth]{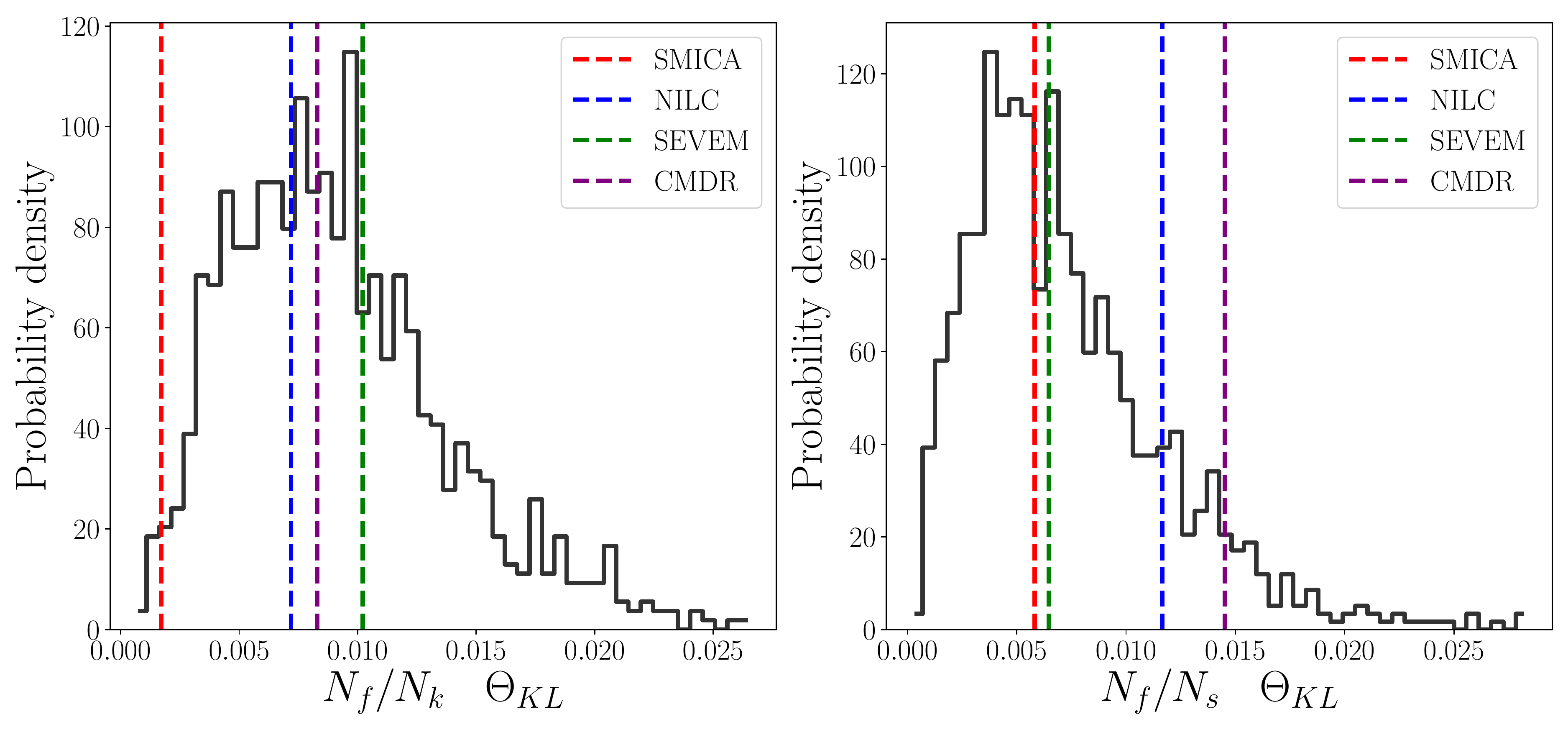}
  \caption{
    The vertical dotted lines are KL-entropy values of the Planck ratio distributions against an average background
    from Gaussian simulations. The black line is the
    distribution of the KL-entropy values of 1035 unique pairings of many Gaussian realizations. These allow
    us to calculate $p$-values for finding entropies of the Planck maps at least that large.
  }
  \label{fig:kl_distribution}
\end{figure*}

The background Gaussian is created by averaging the
distributions from 46 unique Gaussian simulations. To give meaning to the entropies calculated
in that way, a background distribution for them was also realized. The 46 Gaussian distributions
allow us to match up 1035 unique pairs and calculate the KL-entropy of each pairing. A
density distribution of the background entropy values provides the opportunity of finding a
$p$-value for each measured entropy in the Planck maps. These results are visualized in
Fig.~\ref{fig:kl_distribution} and Table~\ref{tab:kl_pvals}.

\begin{table*}[!ht]
  \centering
  \begin{tabular}{ c|c|c|c|c }
    \hline 
    Map name & $f/k$ KL-entropy & $f/k \ p$-value & $f/s$ KL-entropy & $f/s \ p$-value \\ 
    \hline 
    SMICA & 0.0017 & 0.9884 & 0.0058 & 0.5044 \\ 
    NILC & 0.0072 & 0.6135 & 0.0117 & 0.1560 \\ 
    CMDR & 0.0083 & 0.5130 & 0.0145 & 0.0713 \\ 
    SEVEM & 0.0102 & 0.3324 & 0.0065 & 0.4661 \\ 
  \end{tabular}
  \caption{
    Data from all Planck maps. $p$-values give the probability of finding a KL-entropy at least that large.
  }
  \label{tab:kl_pvals}
\end{table*}

\section{Conclusions}
\label{sec:6}

The non-polarized points in the CMB polarization field can be classified according to the
local geometry of the field, which is converged in the points of local minima of the intensity
and anomalies of the polarization angle. 
Gaussianity predicts that the different types of
NPP should occur in certain fixed ratios. By measuring the
variation of these ratios in different subareas of the sky, on can 
construct a sensitive test of non-Gaussianity and statistical anisotropy, complimentary to
\cite{2016,2020}.

We separate the $Q$ and $U$ Stokes parameters into the E and B- families and applied the NPP theory to these families. Our analysis
revealed the regularities of their statistics depending on the amplitude of the
tensor/scalar ratio $r$ and lensing effect. We have shown that in the absence of lensing,
the total number of NPPs of all types does not depend on the $r$ parameter, while
the lensing effect removes this degeneracy. Thus, the analysis of NPP is a powerful tool for
investigating the manifestation of the effect of lensing for the next generation of CMB experiments
devoted to detect or constrain cosmological models with $r\le 10^{-2}$. The same test can be
used for detection of E/B leakage corrections.

We have applied theory of NPP to the Planck 2018 CMB polarization maps and showed, that there is general consistency of SMICA and
NILC compared to a reference set of Gaussian simulations. The strongest
discrepancies are found in Commander ($p \approx 0.07$) and NILC ($p \approx 0.15$).
Although these discrepancies are not highly significant, they may point to the
presence of  foreground, and component separation residuals in
these two maps compared to SMICA and SEVEM. With the upcoming era of high-precision,
ground-based and space-based CMB polarization observations, methods exploiting NPP
statistics will be useful for characterizing and testing the statistical properties of derived maps.

\acknowledgments

The HEALPix pixelization \cite{healpix} scheme was heavily used in this work, and we thank
them for their contributions to the field. This research was partially funded by the Villum
Fonden through the Deep Space project.

\bibliography{article}

\end{document}